\documentclass[twocolumn,showpacs,preprintnumbers,amsmath,amssymb]{revtex4}

\usepackage{graphicx}
\usepackage{dcolumn}
\usepackage{bm}
\usepackage{amsmath,epsfig}
\usepackage{latexsym}

\begin{document}

\title{Correlation based networks of equity returns sampled at different time horizons}

\author{M. Tumminello \dag, T. Di Matteo \ddag, T. Aste \ddag, R.N. Mantegna \dag}

\affiliation{%
\dag Dipartimento di Fisica e Tecnologie Relative, 
Universit\`a di Palermo, Viale delle Scienze, I-90128 Palermo, Italy.\\
\ddag Department of  Applied Mathematics, Australian National University,
0200 Canberra, ACT, Australia.
}%

\date{\today}

\begin{abstract}
We investigate the planar maximally filtered graphs of the portfolio of the 300 most capitalized stocks traded at the New York Stock Exchange during the time period 2001-2003. 
Topological properties such as the average length of shortest paths, the betweenness and the degree are computed on different planar maximally filtered graphs generated by sampling the returns at different time horizons ranging from 5 min up to one trading day.  
This analysis confirms that the selected stocks compose a hierarchical system progressively structuring as the sampling time horizon increases. 
Finally, a cluster formation, associated to economic sectors, is quantitatively investigated.
\end{abstract}


\maketitle

\section{Introduction}

The study of networks is currently a hot topic of research and the topological properties of several graphs describing physical  and social systems have been extensively investigated.  
Early examples are the World Wide Web \cite{Albert99}, correlation based networks in finance \cite{Mantegna99}, Internet  \cite{Faloutsos99,ALBarabasi02,Caldarelli2000,Pastor2001}, and social networks \cite{Wasserman,Newman2002}. 
Other examples include scientific citations \cite{Redner98}, sexual contacts among individuals \cite{Liljeros} and  food webs \cite{Garlaschelli}.
In these systems, it results that the network of links between elements has peculiar topological properties that differ from the ones of a regular or random graph \cite{Erdos,Watts98,Barabasi99}. 
The challenge is to uncover whether there is a relation between the particular properties of such networks and the  special properties of these complex systems. 

In the present study, we are considering correlation based networks obtained by analyzing the price dynamics of a set of stocks simultaneously traded in a financial market. 
This approach generates networks starting from a set of time series. 
Specifically, from a set of $n$ time series one can calculate the correlation coefficient between any pair of variables. 
Each pair of nodes of the network can be thought to be connected by an arc with a weight related to the correlation coefficient between the two time series. 
Such a  network is therefore completely connected.  
By applying a suitable filtering procedure on the network one can remove the less relevant information by
disconnecting some, usually weakly connected elements. 
There are several possible ways of filtering the correlation matrix and the associated network.
We focus on the Planar Maximally Filtered Graph (PMFG) \cite{Tumminello05,Aste05} which is a topological generalization of the  Minimum Spanning Tree (MST) \cite{Mantegna99}. 
MSTs are particular types of graphs that connect all the vertices  through the most correlated link without forming any loop. 
Conversely, the PMFGs are networks containing all the most correlated links which can be kept under the constraint of being representable on a plane without any edge-crossing (planar graph).
It has been shown in Ref.\cite{Tumminello05} that the PMFG always contains the MST as a sub-graph.

The presence of a high degree of cross-correlation between the synchronous time
evolution of a set of equity returns is a well known empirical fact observed in
financial markets \cite{Markowitz59,Elton95,Campbell97}. For a time horizon of one
trading day correlation coefficient as high as 0.7 can be observed for some pair of
equity returns belonging to the same economic sector.
The study of cross-correlation of a set of financial equities has practical
importance since it can improve the ability to model composed financial entities such
as, for example, stock portfolios. There are different approaches to address this
problem. The most common one is the principal component analysis of the correlation
matrix of the data \cite{Elton71}. An investigation of the properties of the
correlation matrix has been performed by physicists by using the perspective and
theoretical results of the random matrix theory \cite{Laloux99,Plerou99}.  As
mentioned above, another approach is the correlation based clustering analysis which
allows to obtain clusters of stocks starting from the time series of price returns.
Different algorithms exist to perform cluster analysis in finance 
\cite{Mantegna99,Tumminello05,Aste05,Panton76,Kullmann2000,Giada2001,Bonanno01,Bernaschi2002,Marsili2002,Onnela02,Bonanno03,Onnela03,Bonanno04,DiMatteo04,Onnela04,DiMatteo05,Coronnello05}.
The effectiveness of single linkage and average linkage clustering in portfolio optimization has been recently investigated in Ref. \cite{Tola06}.

In this paper, we discuss how the correlation structure of a portfolio
of stocks changes when the time horizon of return time series, which are used to compute the
correlation coefficient, is progressively decreased to a short intraday  time scale.
It is known since 1979 \cite{Epps79} that the degree of cross-correlation 
diminishes by reducing the time horizon used to compute 
stock returns \cite{Bonanno01,Toth06}. This phenomenon is sometime 
addressed as ``Epps effect". The existence of this phenomenon 
motivates us to investigate the nature and the properties 
of the network associated to a given financial portfolio
as a function of the time horizon used to record stock
return time series.
 
Ref. \cite{Bonanno01} investigated for the first time correlation based networks obtained from time series sampled at different time horizon and including high-frequency intraday data.
By investigating the topological properties of the MST obtained at different time horizons, it was shown that a clear modification of the hierarchical organization of the set of stocks is detected when one changes the time horizon. The investigation was performed with the set of 100 US highly capitalized stock returns when the time horizon of price returns varied from $d=23400$ s to $d/20$, where $d$
is equal to one trading day at the New York Stock Exchange. The shortest time horizon was chosen in order to statistically ensure that for each stock at least $1$ transaction occurs during  $\Delta t$.

The `Epps effect' implies that the pair correlation decreases by decreasing the time horizon $\Delta t$. In Ref. \cite{Bonanno01}, authors show that this effect is indeed clearly detected. 
Ref.  \cite{Bonanno01} has also shown that the decrease of the correlation between pairs of stocks also affects the nature of the hierarchical organization of stocks. Specifically, Ref.  \cite{Bonanno01} shows that the topology of the network evolves from a star like structure characterizing the network obtained for the shortest sampling time horizon to a much more structured network for the longest time horizon (in the specific case one trading day).

In the present study, we investigate the topological properties of the PMFG. 
We show that  the PMFG is able to retrieve all the results obtained for the MST and it provides a more effective methodology to track the topological changes of the correlation based graph. 

The paper is organized as follows:
Sect.  \ref{methods} discusses the methods used to filter out information from data, whereas Sect. \ref{description} and Sect. \ref{constren} focus on the data analysis. Finally in Sect. \ref{conclusions} we draw our conclusions.    

\section{Correlation based clustering} \label{methods}

Let us here summarize the construction procedures for the MST and PMFG graphs.
The MST is a graph which contains no loops and it connects all the $n$ nodes with the shortest $n - 1$ links. 
The selection of these $n-1$ links is done according to some classic algorithm \cite{Papadimitriou82} and can be summarized as follows: (i) construct an ordered list of pair of stocks $L_{ord}$, by ranking all the possible pairs according to their correlation coefficient $\rho_{ij}$ or distance $d_{ij}=\sqrt{2(1-\rho_{ij})}$. The first pair of $L_{ord}$ has the highest correlation or the shortest distance; (ii) the first pair of $L_{ord}$ gives the first two elements of the MST and the link between them; (iii) the construction of the MST continues by analyzing the list $L_{ord}$. At each successive stage, a pair of elements is selected from $L_{ord}$ and the corresponding link is added to the MST if and only if no loops are generated in the graph after the link insertion.

In Ref. \cite{Gower1969} the procedure briefly sketched above has been shown to provide a MST which is associated to the hierarchical tree of the Single Linkage Clustering Algorithm. In this procedure, at each step, when two elements or one element and a cluster or two clusters $p$ and $q$ merge into a wider single cluster $t$, the distance $d_{tr}$ between the new cluster $t$ and any cluster $r$ is recursively determined by  $d_{tr}=\min \{ d_{pr},d_{qr}\}$. By applying iteratively this procedure, $n-1$ elements of the $n(n-1)/2$ distinct elements of the original correlation coefficient matrix are selected. 

The PMFG has been introduced in two recent papers \cite{Tumminello05,Aste05}. The basic idea is to obtain a graph that retains the same hierarchical properties of the MST, but which is allowing a greater number of links and more complex topological structures than the MST, such as loops and cliques. Such a graph is obtained by relaxing the topological constraint of the described MST construction protocol according to which no loops are allowed in a tree. Specifically, in the PMFG a link can be included in the graph if and only if the graph with the new link included is still planar. A graph is planar if and only if it can be drawn on a plane (infinite in principle) without edge crossings \cite{West}. \\

The first difference between MST and PMFG is about the number of links, which is $n-1$ in the MST and $3(n-2)$ in the PMFG. Furthermore loops and cliques are allowed in the PMFG. A clique of $r$ elements, r-clique, is a subgraph of $r$ elements where each element is linked to each other. Because of the Kuratowski's theorem \cite{West} only 3-cliques and 4-cliques are allowed in the PMFG. The study of 3-cliques and 4-cliques is relevant for understanding the strength of clusters in the system \cite{Tumminello05} as we will see below in an empirical application. We will use this property to investigate the topological changes detected in PMFGs obtained for different sampling time horizons.

The topological properties of different graphs will be investigated by considering several indicators. Specifically, we consider (i) the shortest path $s(i,j)$, which is defined as the minimum number of edges crossed by connecting vertices $i$ and $j$ in the graph; (ii) the degree $k(i)$, which is the number of edges connected to the vertex $i$; (iii) the betweenness $btw(i)$ obtained as the number of shortest paths traversing the vertex $i$ and (iv) the connection strength \cite{Tumminello05}, which is obtained by considering the ratio between the number of cliques of 3 or 4 elements present among $n_s$ stocks belonging to a given set and a normalizing quantity. These normalizing quantities are $n_s-3$ for 4-cliques and $3\,n_s-8$ for 3-cliques \cite{Tumminello05}.  

\section{Empirical Analysis of PMFG networks} \label{description}
We perform our investigation on the 300 most capitalized stocks traded at New York Stock Exchange (NYSE) during the time period January 2001 to December 2003.  
The capitalization value is considered at 12/2003. The return time series are sampled at different time horizons: 5,15,30,65,130,195 and 390 min. The last time horizon of 390 min corresponds to a trading day.

\begin{figure*}
\resizebox{1.8\columnwidth}{!}{\includegraphics{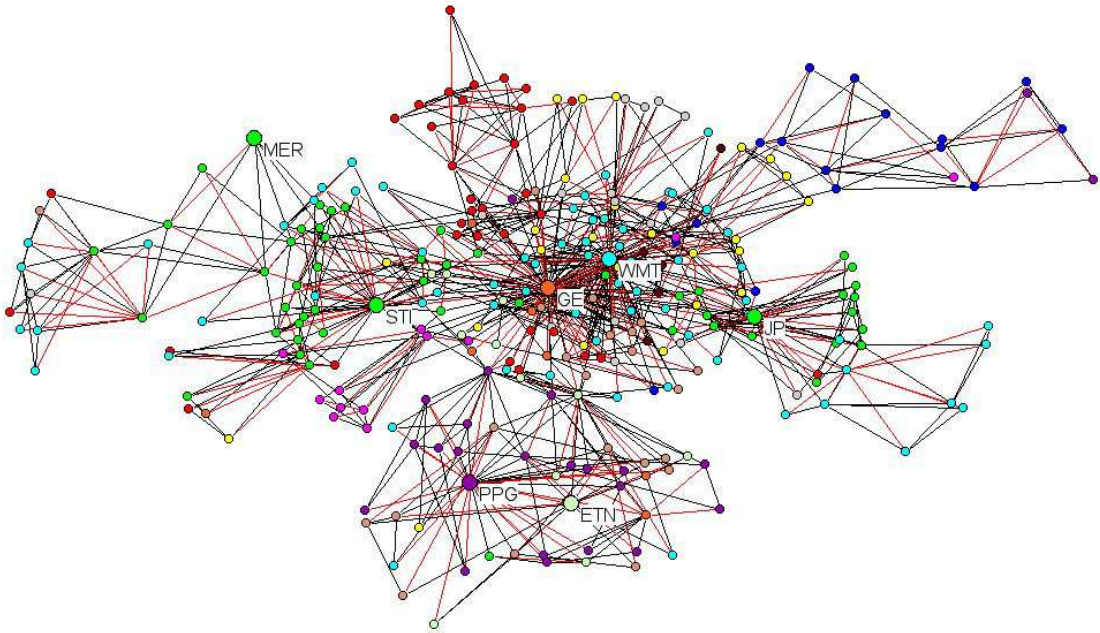}}
\caption{ PMFG at 5 min time horizon. The tick symbol is reported only for 7 stocks, i.e. General Electric (GE), Merrill  Lynch co inc (MER), Wal-Mart stores inc (WMT), Suntrust banks inc (STI), PPG industries inc (PPG), Eaton corp (ETN) and Jefferson-Pilot corp (JP). These stocks are selected by ranking stocks in decreasing order with respect to the degree and picking up stocks in the first 5 ranking positions in at least one of the graphs of Fig. \ref{PMFG5min} and Fig. \ref{PMFGday}. Vertices corresponding to labeled stocks are also highlighted by circles of larger size. Red links in the graph are corresponding to the MST which is contained in the PMFG. Vertices are drawn with different colors to highlight the economic sector each stock belongs to. Specifically these sectors are Basic Materials (violet, 24 stocks), Consumer Cyclical (tan, 22 stocks), Consumer Non Cyclical (yellow, 25 stocks), Energy (blue, 17 stocks), Services (cyan, 69 stocks), Financial (green, 53 stocks), Healthcare (gray, 19 stocks), Technology (red, 34 stocks), Utilities (magenta, 12 stocks), Transportation (brown, 5 stocks), Conglomerates (orange, 8 stocks) and Capital Goods (light green, 12 stocks).} 
\label{PMFG5min}
\end{figure*}

\begin{figure*}
\resizebox{1.8\columnwidth}{!}{\includegraphics{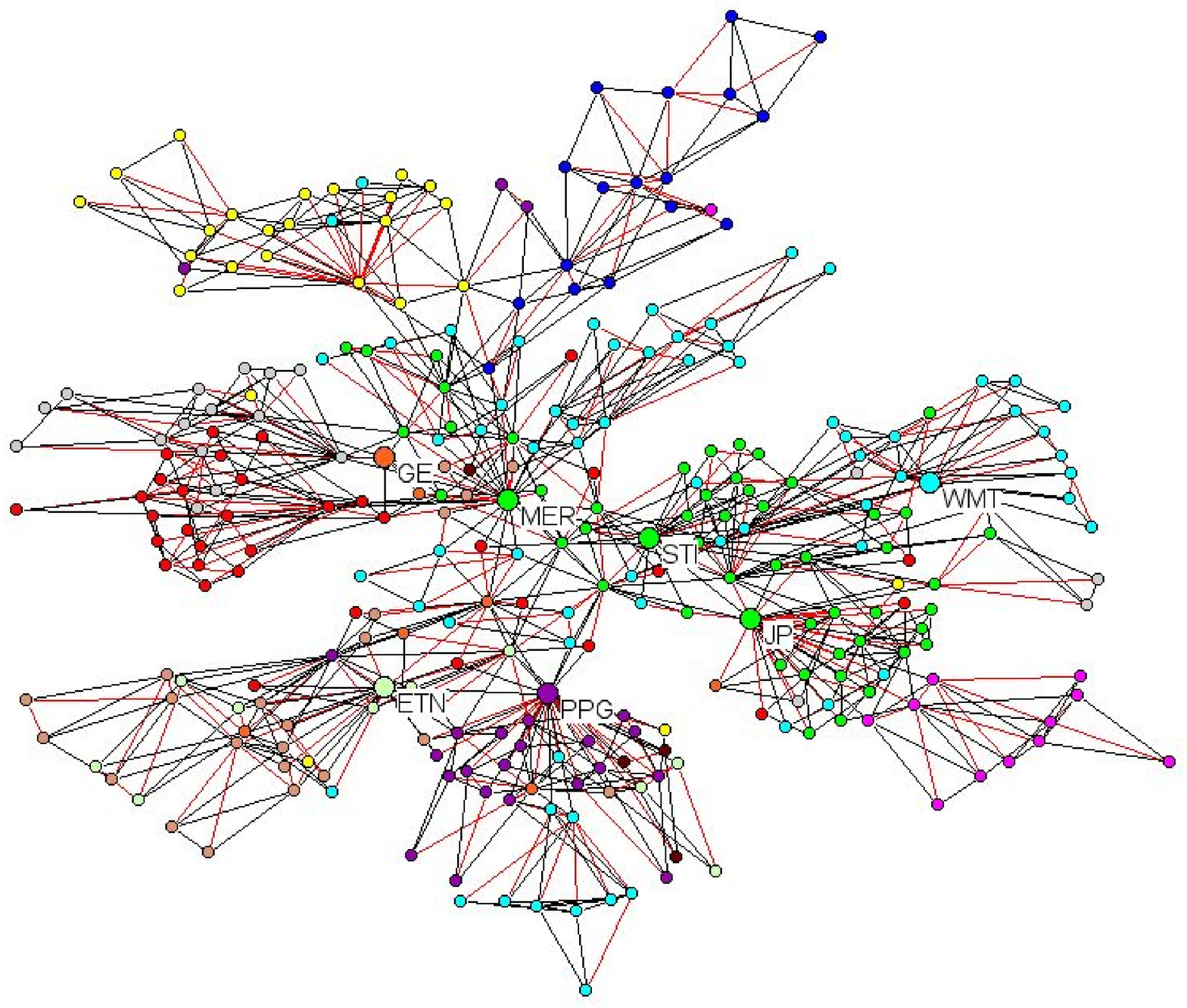}}
\caption{PMFG at daily time horizon. The tick symbol is reported only for 7 stocks, i.e. General Electric (GE), Merrill  Lynch co inc (MER), Wal-Mart stores inc (WMT), Suntrust banks inc (STI), PPG industries inc (PPG), Eaton corp (ETN) and Jefferson-Pilot corp (JP). These stocks are selected by ranking stocks in decreasing order with respect to the degree and picking up stocks in the first 5 ranking positions in at least one of the graphs of Fig. \ref{PMFG5min} and Fig. \ref{PMFGday}. Vertices corresponding to labeled stocks are also highlighted by circles of larger size. Red links in the graph are corresponding to the MST which is contained in the PMFG. Vertices are drawn with different colors to highlight the economic sector each stock belongs to. Specifically these sectors are Basic Materials (violet, 24 stocks), Consumer Cyclical (tan, 22 stocks), Consumer Non Cyclical (yellow, 25 stocks), Energy (blue, 17 stocks), Services (cyan, 69 stocks), Financial (green, 53 stocks), Healthcare (gray, 19 stocks), Technology (red, 34 stocks), Utilities (magenta, 12 stocks), Transportation (brown, 5 stocks), Conglomerates (orange, 8 stocks) and Capital Goods (light green, 12 stocks).} 
\label{PMFGday}
\end{figure*}

In Figs. 1 and 2 we plot the PMFGs computed with the 5 min and 390 min time horizons respectively. In each figure the links also present in the MSTs are drawn in the PMFGs in red colors. For the sake of readability of the pictures, the tick symbol is reported only for 7 stocks. These stocks are selected by ranking stocks in decreasing order with respect to the degree and picking up stocks in the first 5 ranking positions in at least one of the graphs of Fig. \ref{PMFG5min} and Fig. \ref{PMFGday}. The two networks are quite different. In fact the PMFG of 5 min return time series is more compact than the one obtained for the 390 min time series. This last network clearly shows several branches of vertices quite separate from the general framework. Another difference concerns the presence of a few vertices characterized by a very large value of their degree. This is observed clearly in Fig. 1 and it is less evident in Fig. 2. \\

\begin{figure}
\resizebox{1\columnwidth}{!}{\includegraphics{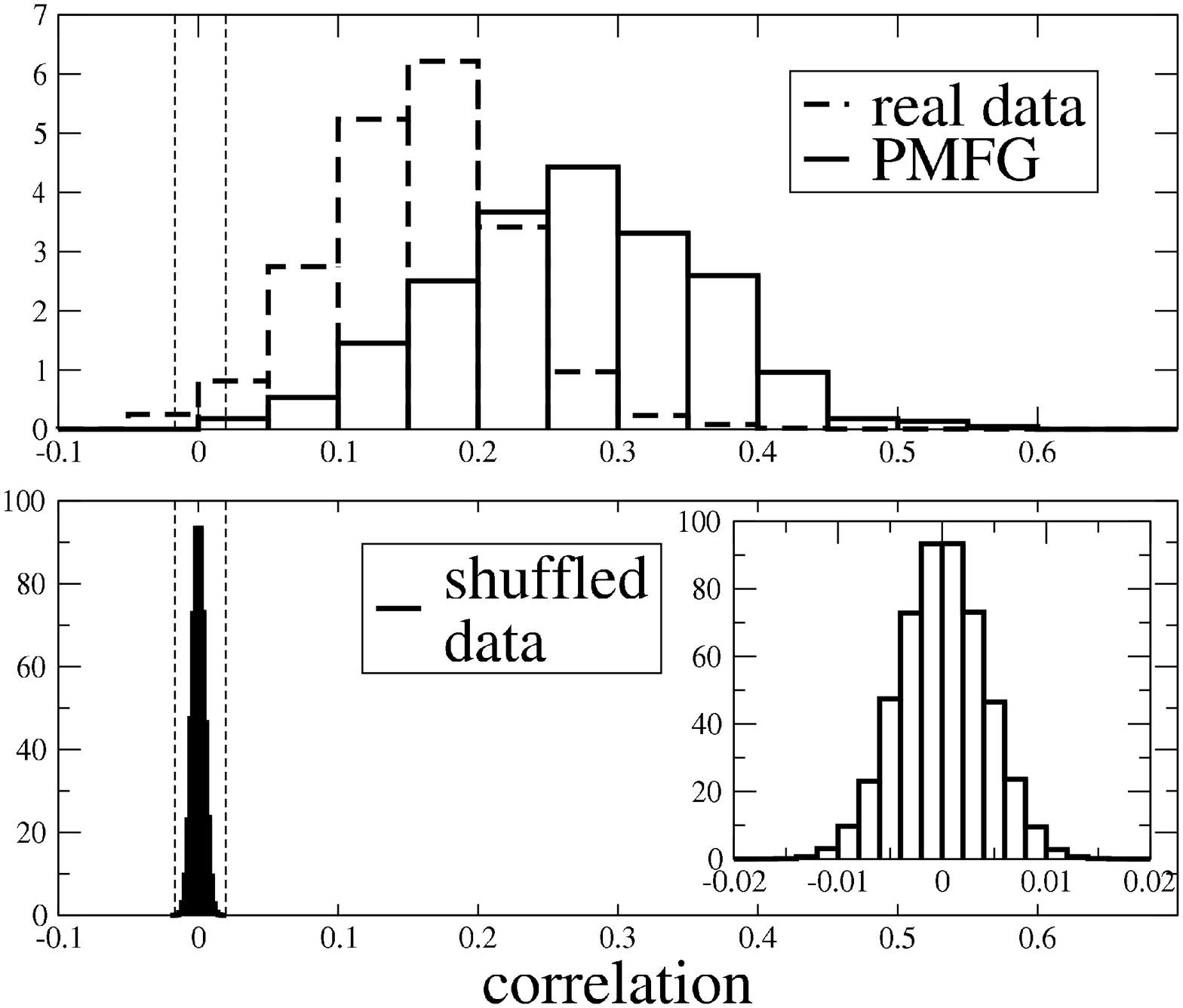}}
\caption{Top panel: comparison between the histogram of correlation coefficients belonging to the empirical correlation matrix estimated at 5 minute time horizon and the corresponding correlation coefficients selected by the PMFG. Bottom panel: histogram of the correlation coefficients obtained by randomly shuffling the 5 minute returns of the 300 stocks. The two  vertical dashed lines correspond to the maximum and minimum value of correlation coefficients obtained in the shuffled data set both in the top and bottom panels.} 
\label{PMFG5minSTAT}
\end{figure}

\begin{figure}
\resizebox{1\columnwidth}{!}{\includegraphics{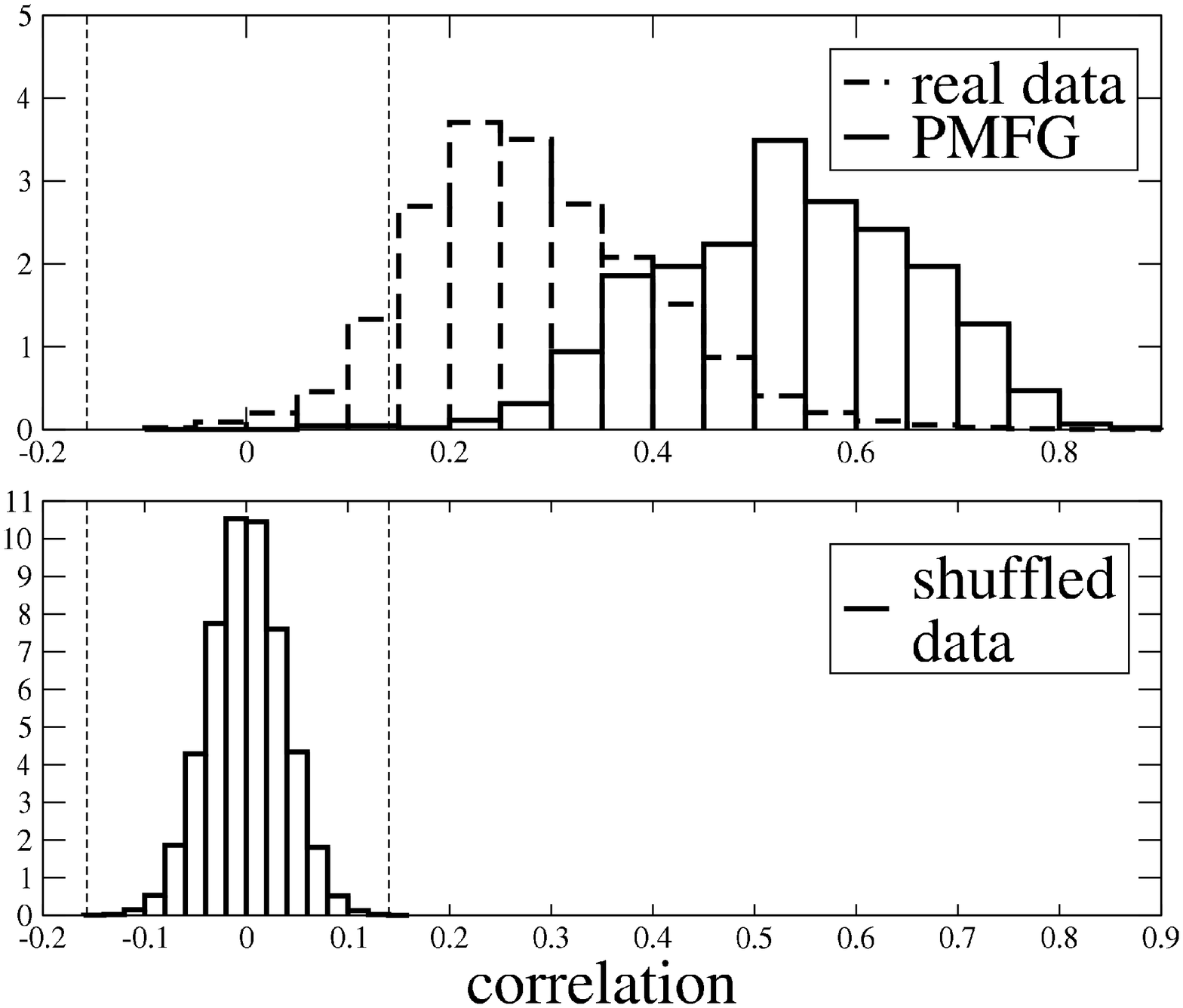}}
\caption{Top panel: comparison between the histogram of correlation coefficients belonging to the empirical correlation matrix estimated at daily time horizon and the corresponding correlation coefficients selected by the PMFG. Bottom panel: histogram of the correlation coefficients obtained by randomly shuffling the daily returns of the 300 stocks. The two vertical dashed lines correspond to the maximum and minimum value of correlation coefficients obtained in the shuffled data set both in the top and bottom panels.} 
\label{PMFGdaySTAT}
\end{figure}

It is certainly relevant to ask at which extent are the graphs carrying information about the system? How robust are they? To answer these questions, in Fig. \ref{PMFG5minSTAT} and Fig. \ref{PMFGdaySTAT} we compare the probability density function (pdf) of correlation coefficients present in the correlation matrix with the pdf of correlations selected by the PMFG (top panel of both the figures) at the 5 minute time horizon and daily time horizon respectively. In the bottom panel of both figures we plot the pdf of correlation coefficients for surrogated multivariate time series obtained by randomly shuffling the return time series of every stock. Figures \ref{PMFG5minSTAT} and \ref{PMFGdaySTAT} show that the PMFG selects correlation coefficients which are in average larger than the average correlation of the empirical correlation matrix. More precisely at 5 minute time horizon the average value of correlation coefficient is 0.16 and the standard deviation of the pdf is 0.07 whereas the average value of correlation coefficient of links selected by the PMFG is 0.27 and its standard deviation is 0.09. At daily time horizon the average correlation coefficient is 0.28 with a standard deviation of 0.12 while the average value of correlation coefficient of links in the PMFG is 0.53 with a standard deviation of 0.12. These results indicate that the PMFG selects most of links among the pair of elements with the highest correlation coefficient. Furthermore a comparison of top panel and bottom panel of Fig. \ref{PMFG5minSTAT} shows that most of the correlation coefficients present in the correlation matrix as well as the correlation coefficient of links selected in the PMFG are not in agreement with the null hypothesis of uncorrelated stock returns. Indeed the average value of correlation coefficient for the shuffled data set is $-7 \times 10^{-6}$ with a standard deviation of 0.004. The minimum and maximum values of correlation obtained by shuffling the 5 minute return time series are -0.017 and 0.020  respectively. It is worth noting that 581 of the $n (n-1)/2=44850$ correlation coefficients have a value belonging to the range $[-0.017,0.020]$ whereas only 5 of the $3 (n-2)=894$ correlation coefficients selected by the PMFG are belonging to the same range at 5 minute time horizon. At the daily time horizon the standard deviation of the correlation coefficients obtained by shuffling the return data is 0.04, i.e. an order of magnitude larger than the corresponding value obtained at 5 minute time horizon. This fact is due to the different number of records of the two time series which is 58344 records for the 5 minute time horizon and 748 records for the daily time horizon. The minimum value of correlation coefficient obtained by shuffling the daily return data set is -0.16 whereas the maximum is 0.14. In the daily return correlation matrix the number of correlation coefficients belonging to the range of values $[-0.16,0.14]$ is 3929 whereas the number of links selected by the PMFG with a correlation coefficient value lying in this range is only 4. Comparing the results obtained at 5 minute time horizon with those obtained at daily time horizon we notice that the percentage of links in the PMFG that are in agreement with the null hypothesis of uncorrelated data is for both time horizons of the order of $0.5 \%$ whereas the percentage of correlation coefficients that are in agreement with the null hypothesis of uncorrelated data is rather different: $1.3 \%$ at the 5 minute time horizon and $8.8 \%$ at the daily time horizon. On the basis of these results, we conclude that the PMFG is carrying information about the strongest interactions observed in the system and it is disregarding most of the correlations consistent with the null hypothesis of uncorrelated data. \\
How robust are these graphs with respect to the statistical uncertainty? At a first glance one might be tempted to answer this question by saying that graphs are robust because we have shown above that most of the correlation coefficients selected by the PMFG are large enough to be considered significant. However in ref. \cite{Tumminello2007} it has been shown that the strength of the correlation of links is a rough measure of their statistical robustness for the MST. To investigate the statistical robustness of links selected by the PMFG we apply the technique proposed in \cite{Tumminello2007}, i.e. we construct 1000 bootstrap replicas of data and from every replica data set we extract the PMFG. In the following we shall refer to such graphs obtained by bootstrap replicas as PMFG$^*(r)$ with $r=1,...,1000$. We associate to each link of the PMFG of the empirical data a number called bootstrap value and corresponding to the percentage of PMFG$^*$s in which the considered link of the PMFG appears. The histogram of bootstrap values obtained for both the 5 minute time horizon and the daily time horizon shows a prominent pick for bootstrap values larger than $95\%$. At 5 minute time horizon the number of links with a bootstrap value larger than $95\%$ is 353/894 whereas at daily time horizon is 190/894. This result suggests that the PMFG describing the system at 5 minute time horizon is statistically more robust than the PMFG of daily returns. This fact can be interpreted as the consequence of two effects: 1) the different number of records of time series, i.e. 58344 for the 5 minute time horizon and 748 for the daily time horizon; 2) the different level of complexity of the system at different time horizon. Indeed the structure of PMFG for 5 minute return reveals a star-like shape which is a topologically simple and robust structure whereas at daily time horizon the structure is rather complex.\\

\begin{figure}[h]
\resizebox{1\columnwidth}{!}{%
  \includegraphics{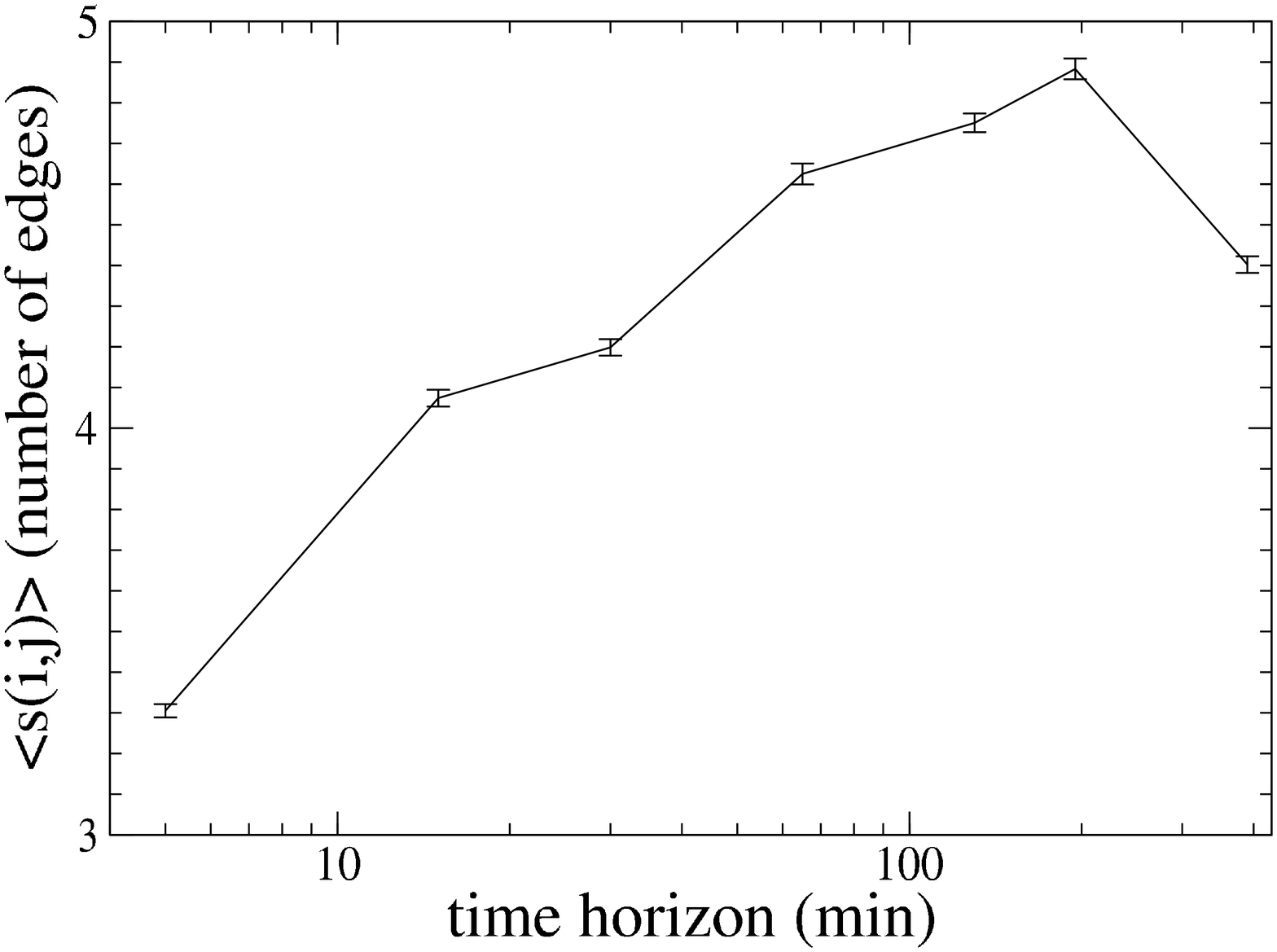}}
\caption{Average length of shortest path in the PMFG as function of the sampling time horizon of return time series.} 
\label{shortpath}
\end{figure}

We quantify the degree of compactness of the PMFGs by computing the average length of shortest paths at different time horizons. The result of our investigation is shown in Fig. \ref{shortpath}. Specifically, we observe that the average length of shortest paths is about 3.3 when the time horizon is 5 min. The average length increases by increasing the time horizon and reaches the maximum value of 4.9 for the time horizon of 195 min and then decreases to the value 4.4 observed for the daily (390 min) time horizon. We therefore observe a progressive structuring of the PMFGs as a function of the time horizon with a maximal level of structuring present for $\Delta t=195$ min. We have not yet an explanation for the presence of a maximal value of the  average length occuring at an intraday time horizon.

\begin{figure}[t]
\resizebox{1\columnwidth}{!}{%
  \includegraphics{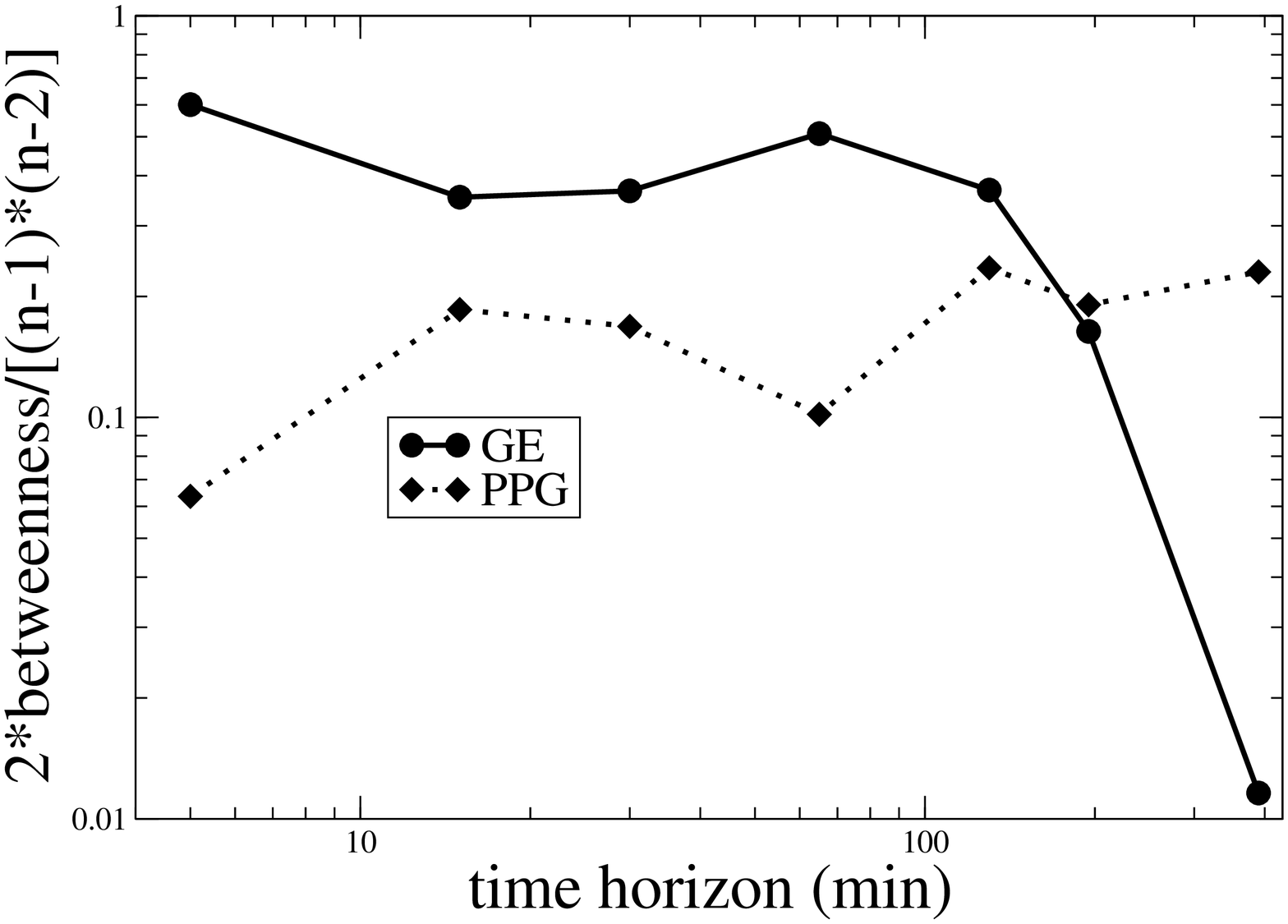}}
\caption{Betweenness of GE and PPG evaluated in the PMFG as function of the time horizon. The maximal betwenness of the PMFG is observed for GE when $\Delta t \le 130$ min and for PPG when $\Delta t > 130$ min.} 
\label{betweennessGE}
\end{figure}

\begin{figure}[t]
\resizebox{1\columnwidth}{!}{%
  \includegraphics{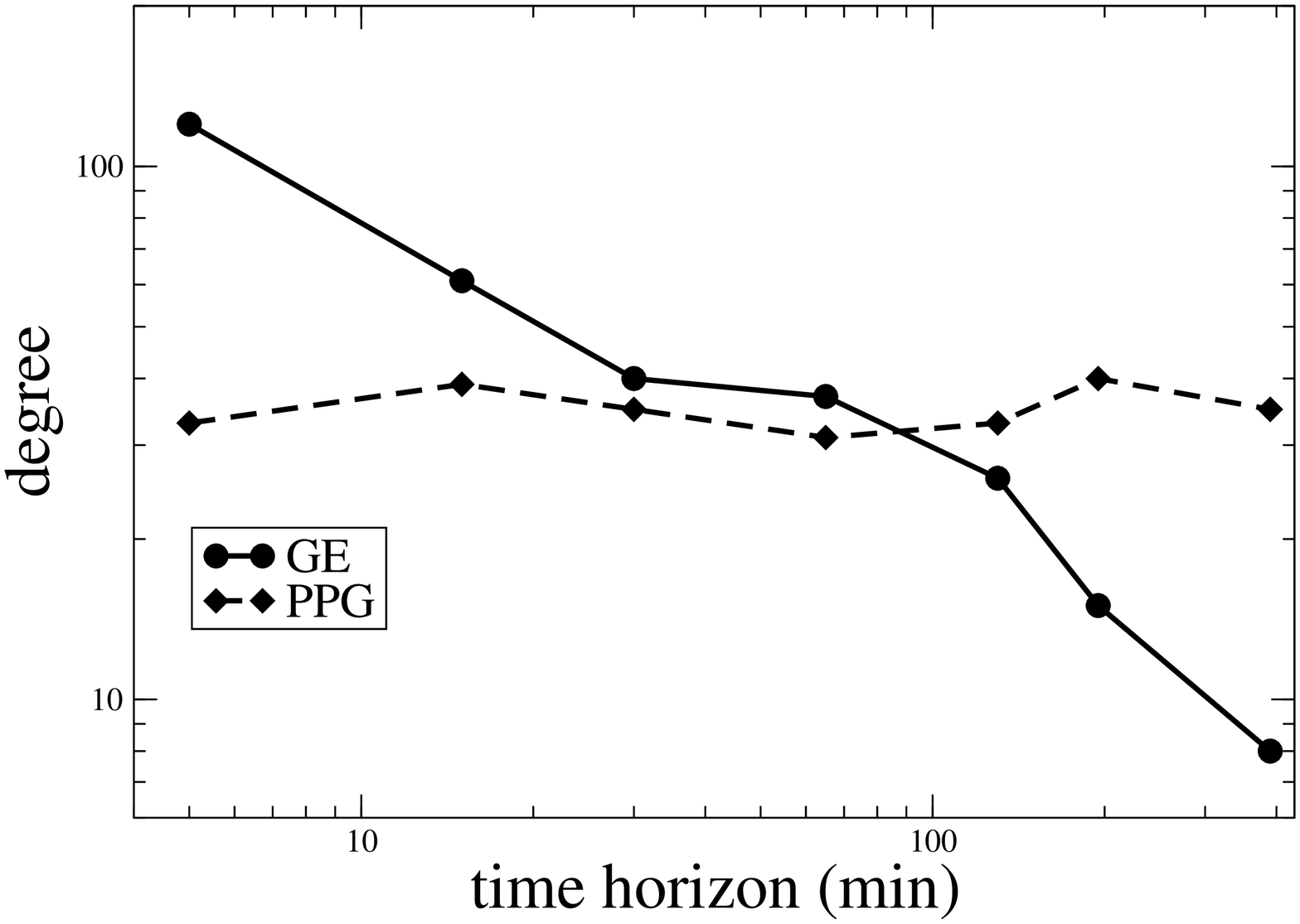}}
\caption{Degree of GE and PPG evaluated in the PMFG as function of the time horizon. The maximal degree of the PMFG is observed for GE when $\Delta t \le 65$ min and for PPG when $\Delta t > 65$ min.} 
\label{GEdegree}
\end{figure}

We characterize the topological properties of PMFGs obtained at different time horizons by also measuring the maximal betweenness and the maximal degree of networks. In Fig. \ref{betweennessGE} and in Fig. \ref{GEdegree} we show the betweenness and degree of the two stocks that assume the maximal value of the betweenness and of degree at different time horizons. Specifically, the maximal betweenness is detected for General Electric (GE) at short time horizons and for PPG Industries at time horizons equal or longer than 195 min. A similar alternate profile is also observed for the degree in Fig. \ref{GEdegree}. In fact, for short time horizons the stock with the maximal degree is GE whereas starting from $\Delta t=130$ min the stock with the highest degree is PPG. It should also be noted from the figures that both the betweenness and the degree for GE is monotonously decreasing when the time horizon increases whereas the corresponding values observed for PPG are roughly constant. The crossover of the two profiles approximately occurs for time horizon close to the interval 130-195 min. This time interval contains the time horizon where the maximum of the average length of shortest paths is detected. It should also be noted the different role that GE and PPG play in the system. The stock GE is a hub for the whole market whereas the stock PPG is a hub for its own economic sector (Basic Materials). Following this reasoning one can interpret results of Fig. \ref{betweennessGE} and Fig. \ref{GEdegree} in terms of the different interaction between economic sectors at different time horizons. Indeed the internal structure of sectors is roughly formed already at short time horizons (see Sect. \ref{constren}) and this explains the behavior of betweenness and degree of PPG as a function of the time horizon. Furthermore Fig. \ref{PMFG5min} and Fig. \ref{shortpath} suggest that GE at short time horizons strongly intervenes in the connection between different branches (sectors) of the PMFG whereas at longer time horizons connections between sectors are more complex and the central role of GE progressively disappears. 

\section{Connection strength of economic sectors on PMFGs}\label{constren}

\begin{figure*}[t]
\resizebox{1.9\columnwidth}{!}{\includegraphics{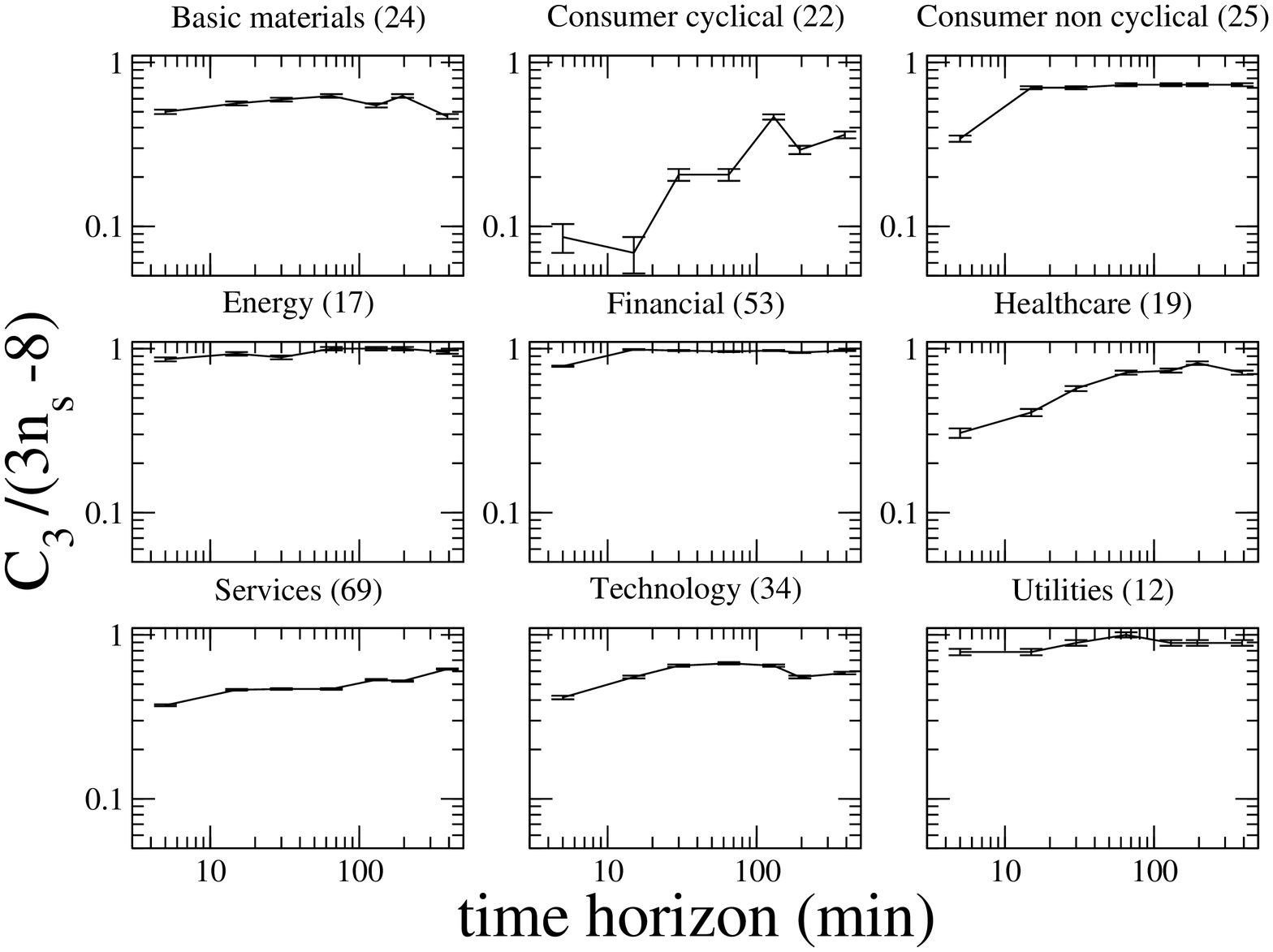}}
\caption{ Connection strength evaluated by the number of intra-sector 3-cliques ($C_3$). 
Error bars are accounting for digitalization error: $1/(3 n_s - 8)$ where $n_s$ is the number of stocks belonging to the sector $s$.} 
\label{ConStren_c3}
\end{figure*}

The topological changes detected in PMFGs have also been investigated by monitoring the connection strength inside specific sectors at different time horizons. Economic sectors of stocks are defined by using the Yahoo finance classification of stocks (April 2005). We quantify the connection strength of a subgroup of $n_s$ elements by counting the number $C_3$ of 3-cliques formed by the elements of the group divided by $3n_s -8$, which is the number of potential 3-cliques that can be formed by $n_s$ elements in a planar graph. A detailed description of this method and of its usefulness is given in Ref. \cite{Tumminello05}. The results of our investigation are summarized in Fig. \ref{ConStren_c3}. The connection strength is shown as a function of the time horizon. In Fig. \ref{ConStren_c3} we investigate 9 sectors comprising 275 stocks over a total of 300. These economic sectors are basic materials (24 stocks), consumer cyclical (22 stocks), consumer non cyclical (25 stocks), energy (17 stocks), financial (53 stocks), healthcare (19 stocks), services (69 stocks), technology (34 stocks) and utilities (12 stocks). The sectors of conglomerates and capital goods are not considered in the figure because they have a connection strength low value, which is almost independent of the specific time horizon.\\ 
In Fig. \ref{ConStren_c3} each panel refers to a single sector.    
The nine panels of Fig. \ref{ConStren_c3} show a variety of behavior. Specifically, there are sectors like energy, financial and utilities where the connection strength is very close to one already at the shortest time horizon. 
This behavior indicates that the sectors are well defined and driven by the same factors down to a very short time horizon. On the other hand, there are sectors like consumer cyclical,  healthcare and services clearly showing that the market needs a finite time to produce a profile of correlation compatible with the sector classification.  Moreover, the value of the connection strength observed for these sectors is always smaller than 1 at longer time horizons. This fact indicates that the PMFG analysis does not interpret the stocks of the considered sector as belonging to a compact region of the PMFG. This might be due to a failure of the filtering ability of PMFG or might just reflect a marked heterogeneity in the classification methods used in defining the economic sectors. Basic materials, consumer non cyclical, and technology sectors show an intermediate behavior characterized by a non marked time dependence and moderately low values of the overall connection strength.   

In order to test whether the behavior observed for different sectors has an economic motivation, therefore supporting the hypothesis that PMFG is able to detect proper communities in the correlation based networks, we have investigated some sub-sectors.
Interesting cases are one sub-sector of the energy economic sector i.e. the oil well services \& equipment sub-sector (5 stocks) and one sub-sector of the utilities sector, i.e. the electric utilities sub sector (10 stocks). Both these sub-sectors have a connection strength equal to 1 and constant over the spanned time horizon. A different behavior is observed for the sub-sector major drugs of the healthcare economic sector, which is composed by 7 stocks and it has associated a connection strength of $0.85 \pm 0.08$ at $T=5$ min and $T=15$ min and connection strength 1 for $T \ge 30$ min. Finally, the sub-sector food processing (consumer non cyclical) is formed by 11 stocks and has a connection strength time evolving profile as summarized here: $(T= 5$ min, $0.56\pm0.04)$, $(T=15$ min, $0.8\pm0.04)$, $(T=30$ min, $0.8\pm0.04)$, $(T=65$ min, $0.92\pm0.04)$, $(T=130$ min, $0.92\pm0.04)$, $(T=195$ min, $0.76\pm0.04)$, $(T=390$ min, $0.92\pm0.04)$. In summary all the considered sub-sectors show a connection strength greater or at most equal to the connection strength of the economic sector they belong to. Furthermore all the considered sub-sectors are significantly intra-connected before or at most at the same time horizon as the corresponding economic sector.\\
 
\section{Conclusions}\label{conclusions}
The measure of the average length of shortest path in the PMFG shows a small world effect present in the networks at any time horizon. The amount of the effect is varying with the sampling time horizon. The more structured network is observed for the intraday time horizon of 195 min. The study of the degree and the betweenness of stocks allows to distinguish the market role of representative stocks. For example, GE is a hub for the whole market at short time horizons and its relevance decreases according to the structuring of the market into sectors observed at long time horizon. 
On the contrary, the stock PPG results to be a local hub (i.e. hub of the economic sector it belongs to - basic materials) both at short and long time horizons. The fact that the betweenness and the degree of PPG  are roughly constant as a function of the time horizon suggests that the sector of basic materials is formed already at short time horizons. This fact is also supported by the measure of connection strength for the specific considered sector. Other sectors, such as consumer cyclical, healthcare and services are characterized by a different behavior showing a time dependence in the structuring of the stocks belonging to these sectors. Finally, we have observed economic sub-sectors such as oil well services \& equipment and  electric utilities being already formed at the 5 min time horizon and characterized by the maximum allowed value of the connection strength, i.e. 1. These results show that the market is progressively structured as a function of the time horizon. The analysis of connection strength of sub-sectors and sectors as a function of the time horizon suggests that the market structuring occurs by first connecting stocks belonging to the same sub-sector and then connecting stocks belonging to the same economic sector. In the present study, the analysis of market structuring has been done by using the classifications of economic sectors provided by Yahoo finance. It would be interesting to use an unsupervised approach and see how homogenous stock communities emerge in the PMFG as a function of the time horizon. Unfortunately, an unsupervised identification of communities or clusters in the PMFG cannot be done by straightforwardly applying techniques of graph theory as the ones discussed for example in ref. \cite{Newman04a,Newman04b,Newman06a}. The reason is that these techniques have been developed to be applied to networks without any specific topological constraint. Both the MST and PMFG are constructed by imposing topological constraints to be satisfied. We therefore think that an appropriate modification of the mentioned methods is needed to perform an unsupervised identification of communities in these graphs.

{\bf Acknowledgments }\\
MT and RNM wish to thank Fabrizio Lillo and Claudia Coronnello for fruitful discussions.
Authors acknowledge partial support from  COST P10 ``Physics of Risk'' project and MIUR 449/97 project ``Ultra-high frequency dynamics of financial markets''.
TDM and TA wish to thank  the partial support by ARC Discovery Projects: DP03440044 (2003) and DP0558183 (2005).
MT and RNM acknowledge partial support from the research project MIUR-FIRB RBNE01CW3M ``Cellular Self-Organizing nets and chaotic nonlinear dynamics to model and control complex systems" and from the European Union STREP project n. 012911 ``Human behavior through dynamics of complex social networks: an interdisciplinary approach".


\end{document}